\documentstyle[12pt]{article}

\def\+{{+\!\!\!+}}

\def\d{\partial}

\def\pmb#1{\setbox0=\hbox{#1}%
\kern.0em\copy0\kern-\wd0
\kern-.04em\copy0\kern-\wd0
\kern.08em\copy0\kern-\wd0
\kern-.04em\raise.0433em\box0 }         

\newcommand{\nc}{\newcommand}
\nc{\beq}{\begin{equation}}
\nc{\eeq}[1]{\label{#1}\end{equation}}
\nc{\ber}{\begin{eqnarray}}
\nc{\eer}[1]{\label{#1}\end{eqnarray}}
\nc{\pek}[1]{\cite{#1}}
\nc{\enr}[1]{(\ref{#1})}
\nc{\kal}[1]{{\cal{#1}}}
\nc{\dott}{\;\cdot\;}

\newcommand{\be}{\begin{equation}}
\newcommand{\ee}{\end{equation}}
\newcommand{\bea}{\begin{eqnarray}}
\newcommand{\eea}{\end{eqnarray}}

\begin{document}
\begin{center}
                                \hfill   hep-th/0405240\\
                                \hfill   LPTHE-04-10\\

\vskip .3in \noindent

\vskip .1in

{\large \bf Geometry of D-branes for general $N=(2,2)$ sigma models}
\vskip .2in

{ \bf Maxim Zabzine}\footnote{e-mail address: zabzine@lpthe.jussieu.fr} \\


\vskip .15in

\vskip .15in
{\em Lab. de Physique Th\'eorique et Hautes Energies}\\
{\em Universit\'e Pierre et Marie Curie, Paris VI}\\
{\em  4 Place Jussieu, \ 75252 Paris Cedex 05, France}
\vskip .15in
{\em Institut Mittag-Leffler\\
Aurav\"agen 17,
S-182 62 Djursholm, Sweden}

\bigskip


 \vskip .1in
\end{center}
\vskip .4in

\begin{center} {\bf ABSTRACT }
\end{center}
\begin{quotation}\noindent
 We give a world-sheet description of D-brane in terms of gluing conditions on $T{\cal M} \oplus T^*{\cal M}$. 
 Using the notion of generalized K\"ahler geometry we show that A- and B-types D-branes
 for the general $N=(2,2)$ supersymmetric sigma model (including a non-trivial NS-flux)
 correspond to the (twisted) generalized complex submanifolds with respect to the different
 (twisted) generalized complex structures however.
\end{quotation}
\vfill
\eject


\section{Introduction}

The aim of this note is to give a geometrical description of A- and B- types of branes for the {\em general}
 $N=(2,2)$ sigma model. These branes correspond to a maximally supersymmetric boundary conditions for the 
 $N=(2,2)$ sigma model, i.e. they preserve the half of the bulk world-sheet supersymmetry. 

 Despite the fact that these sigma models were known for twenty years \cite{Gates:1984nk} 
 the appropriate  geometrical language has been missing. Recently Hitchin introduced the notion of generalized
  complex geometry \cite{Hitchin} which provides a natural set up for the relevant geometry. Later the geometry 
 of  the {\em general} $N=(2,2)$ sigma model has been formalized in Gualtieri's thesis \cite{Gualtieri} and now it goes 
 under the name of (twisted) generalized K\"ahler geometry.
 The algebraic set-up for the generalized K\"ahler geometry has been discussed previously in physics 
 literatures \cite{Kapustin:2000aa}. 

This paper is inspired by Kapustin's work \cite{Kapustin:2003sg} and it presents
 the generalization of some of his proposals.  
 The idea is to analyze the geometry of a sigma model and its D-branes not in terms of tangent 
 space $T{\cal M}$ but in terms of tangent and cotangent  spaces $T{\cal M}\oplus T^*{\cal M}$.
 We will show that the appropriate D-branes have a simple description in terms of $T{\cal M}\oplus T^*{\cal M}$.

The article is organized as follows. In Section \ref{s:geometry} we review the geometry of $N=(2,2)$ general 
 sigma model. In particular we briefly sketch the description of this geometry in terms of $T{\cal M}\oplus T^*{\cal M}$, 
 a generalized K\"ahler geometry recently introduced by Gualtieri \cite{Gualtieri}. Section \ref{s:N1branes} deals with
 the description of $N=1$ boundary conditions in terms of gluing map defined over $T{\cal M} \oplus T^*{\cal M}$.      
  In the following Section \ref{s:N2branes} we discuss the geometrical interpretation of A- and B-types branes
 for the general $N=2$ sigma model. 
  We present some examples of $N=(2,2)$ branes.
Finally, in Section \ref{s:end} we give a summary of the paper with a discussion of open 
 problems.

\section{Generalized K\"ahler geometry}
\label{s:geometry}

In this section we review the results on the general $N=(2,2)$ supersymmetric sigma model
 discovered in \cite{Gates:1984nk}. In particular we discuss the bi-Poisson property of 
 this geometry \cite{Lyakhovich:2002kc}. Finally we sketch the notion of (twisted) generalized
 complex  and generalized K\"ahler geometries  \cite{Gualtieri}.
 This Section will allow us to introduce the notation and some relevant concepts.

 Let us start from the general $N=(1,1)$ sigma model (without boundaries) which is written
  in $N=(1,1)$ superfields
\beq
 S= \int d^2\sigma\,d^2\theta\,\,D_+\Phi^\mu D_- \Phi^\nu (g_{\mu\nu}(\Phi)
 + B_{\mu\nu}(\Phi)) ,
\eeq{actionB}
where $H=dB$ on some patch. Although $B$ is used to write the action (\ref{actionB}) down 
 the theory depends only on the closed three form $H$. Thus a manifold $({\cal M}, g, H)$ with a
 metric and a closed three form admits $N=(1,1)$ supersymmetric sigma model.

 We look for additional supersymmetry transformations of the form
\beq
\delta \Phi^\mu =\epsilon^+ D_+ \Phi^\nu J^\mu_{+\nu}(\Phi)
  + \epsilon^- D_- \Phi^\nu J^\mu_{-\nu}(\Phi) .
\eeq{secsupfl}
 Classically the ansatz (\ref{secsupfl}) is unique for dimensional reasons.
It turns out that the action (\ref{actionB}) is invariant under the transformations
 (\ref{secsupfl}) provided that
\beq
 g_{\nu\mu} J_{\pm\rho}^\mu  = - g_{\rho\mu} J^\mu_{\pm\nu}
\eeq{bla}
 and
\beq
 \nabla^{(\pm)}_\rho J^\mu_{\pm\nu} \equiv J^\mu_{\pm\nu,\rho} +
 \Gamma^{\pm\mu}_{\,\,\rho\sigma} J^\sigma_{\pm\nu} - \Gamma^{\pm\sigma}_{\,\,\rho\nu}
 J^\mu_{\pm\sigma}=0 ,
\eeq{nablJH}
 where one defines two affine connections
\beq
 \Gamma^{\pm\mu}_{\,\,\rho\nu} = \Gamma^{\mu}_{\,\,\rho\nu} \pm g^{\mu\sigma} H_{\sigma\rho\nu}.
\eeq{defaffcon}
  Next  we  have to require the standard on-shell $N=(2,2)$ supersymmetry algebra, i.e. the manifest
 supersymmetry transformations and the second supersymmetry transformations (\ref{secsupfl})
 commute and the commutator of two second supersymmetry transformations (\ref{secsupfl}) gives a translation.
  Thus the supersymmetry algebra requires
 that $J_{\pm}$ correspond to two complex structures.

 This is the full description of the most general N=(2,2) sigma model (\ref{actionB}). Thus the target
 manifold should be a bihermitian complex manifold (i.e., there are two complex structures
 and a metric is Hermitian with respect to both) and the two complex structures should
 be covariantly constant, with respect to the different connections however. Thus a manifold with this data
 $({\cal M}, g, J_+, J_-, H)$ admits $N=(2,2)$ supersymmetric sigma model. If $H=0$ then $J_+$ and $J_-$ give rise
 to the standard K\"ahler structures which maybe different, $J_+ \neq J_-$. 
  
The above geometry can be reformulated in different but an equivalent language. For example, we
 can describe the geometry as being bi-Poisson \cite{Lyakhovich:2002kc}.
  Introducing the two-forms $\omega_\pm = g J_\pm$
 we can show that $(\omega_+^{-1} \pm \omega_-^{-1})$ are Poisson tensors. Infact the bi-Poisson 
 property is equivalent to the conditions (\ref{nablJH}). However the property $dH=0$ is lacking in 
 this description.   

Recently Hitchin \cite{Hitchin} has introduced the notion of generalized complex geometry. 
 In Hitchin's construction 
  $T{\cal M}$ is replaced by $T{\cal M}\oplus T^*{\cal M}$ and the Lie bracket is replaced
 by the appropriate bracket on $T{\cal M}\oplus T^*{\cal M}$, the so called Courant bracket. 
 Thus a generalized complex structure  is an almost 
 complex structure ${\cal J}$ on $T{\cal M} \oplus T^*{\cal M}$ whose $+i$-eigenbundle is Courant involutive.
 Using the language of generalized complex geometry Gualtieri reformulated the geometry of $N=(2,2)$ sigma 
 models and it goes under the name of generalized K\"ahler geometry. This reformulation is essential for
 our further discussion.
Below we will briefly review the 
 relevant concepts of this construction.
 A detailed presentation of
 generalized complex and generalized K\"ahler geometries can be found in Gualtieri's thesis \cite{Gualtieri}. 

 We start by providing the definition of generalized complex structure.
  On $T{\cal M}\oplus T^*{\cal M}$ there is a natural indefinite metric defined by 
 $(X+\xi, X+\xi ) =  i_X \xi$.
In the coordinate basis $(\d_\mu, dx^\mu)$ we can write this metric as follows
\beq
 {\cal I} = \left ( \begin{array}{ll}
                              0 & 1_d \\
                              1_d & 0 
                        \end{array} \right ) .
\eeq{definindefmetric}
 An almost generalized complex structure is a map ${\cal J}: 
 T{\cal M} \oplus T^*{\cal M} \rightarrow T{\cal M} \oplus T^*{\cal M}$
 such that ${\cal J}^2 = -1_{2d}$ and that ${\cal I}$ is hermitian with respect to ${\cal J}$, ${\cal J}^t {\cal I}
 {\cal J} ={\cal I}$. On $T{\cal M}\oplus T^*{\cal M}$ there is a Courant bracket which is defined as follows
\beq
 [X +\xi, Y +\eta ]_c = [X, Y] + {\cal L}_X \eta - {\cal L}_Y \xi -\frac{1}{2} d(i_X \eta - i_Y\xi),
\eeq{defCourbrak}
 where $[\,\,,\,\,]$ is a Lie bracket on $T{\cal M}$.
 This bracket is skew-symmetric but in general does not satisfy the Jacobi identity. 
 However if there is a sub-bundle $L \subset T{\cal M}\oplus T^*{\cal M}$ which is involutive (closed under the Courant bracket) 
 and isotropic with respect to ${\cal I}$ then the Courant bracket on the sections of $L$ does satisfy 
 the Jacobi identity. This is a reason for imposing hermiticity of ${\cal I}$ with respect to ${\cal J}$.
 We can construct the projectors on $T{\cal M} \oplus T^*{\cal M}$
\beq
 \Pi_{\pm} = \frac{1}{2} ( I \pm i {\cal J}),
\eeq{defproj}
 the almost generalized complex structure ${\cal J}$ is integrable if 
\beq
  \Pi_{\mp} [\Pi_{\pm}(X+\xi), \Pi_{\pm}(Y+\eta)]_c = 0,
\eeq{inegrablproj} 
 for any $(X+\xi), (Y+\eta) \in T{\cal M} \oplus T^*{\cal M}$. 
 The Courant bracket on $T{\cal M}\oplus T^*{\cal M}$ can be twisted by a closed three form $H$. Namely given a closed
 three form $H$ one can define another bracket on $T{\cal M}\oplus T^*{\cal M}$ by
\beq
 [X +\xi, Y +\eta ]_H= [X +\xi, Y +\eta ]_c + i_X i_Y H.
\eeq{deftwCorbr} 
 This bracket has similar properties to the Courant bracket. 
 Again if a sub-bundle $L \subset T\oplus T^*$ is closed under the twisted Courant bracket 
 and isotropic with respect to ${\cal I}$, then the Courant bracket on the sections of $L$ does satisfy 
 the Jacobi identity.
Thus in the integrability condition (\ref{inegrablproj}) the Courant 
 bracket $[\,\,,\,\,]_c$ can be replaced by the new twisted
 Courant bracket $[\,\,,\,\,]_H$. 
In local coordinates the integrability conditions have been worked out in \cite{Lindstrom:2004iw}. 

 One important feature  of the (twisted) Courant bracket is that, unlike the Lie bracket, this bracket has a nontrivial 
 automorphism defined by a closed two-form $b$. Namely if we define the following action on $T{\cal M} \oplus T^*{\cal M}$
\beq
 e^b(X + \xi)= X+\xi + i_X b.
\eeq{definaurom}
 with an arbitrary two-form $b$. To the transformation (\ref{definaurom}) we refer as a b-transform.
Under the b-transform the (twisted) Courant bracket behaves as  
\beq
 [e^b(X +\xi), e^b(Y +\eta) ]_H = e^b[X +\xi, Y +\eta ]_{H+db} .
\eeq{autmocourbracket} 
 Thus if $db=0$ the action (\ref{definaurom}) is automorphism of (twisted) Courant bracket.

Another important property is that under the natural projection $p: T{\cal M}\oplus T^*{\cal M}\rightarrow T{\cal M}$
 the (twisted) Courant bracket satisfies the following condition
\beq
 p [ X+ \xi, Y +\eta ]_H = [p(X+\xi), p(Y+\eta)]_H = [X, Y].
\eeq{imprpsbcjna}
 Therefore if a sub-bundle $L$ of $T{\cal M} \oplus T^*{\cal M}$ is closed under the (twisted) Courant bracket then 
 its image $p(L)$ is closed under the Lie bracket and thus corresponds to a foliation. 

Next we present the definition of generalized K\"ahler manifold \cite{Gualtieri}. The (twisted) generalized K\"ahler 
 structure is a pair $({\cal J}_1, {\cal J}_2 )$ of commuting (twisted) generalized complex structures
 such that $G= - {\cal J}_1 {\cal J}_2$ is a positive definite metric on $T{\cal M} \oplus T^*{\cal M}$.
 As it has been shown in \cite{Gualtieri} the generalized K\"ahler geometry $({\cal M}, {\cal J}_1, {\cal J}_2)$
 is completely  equivalent to the geometry of $N=(2,2)$ sigma models $({\cal M}, g, J_+, J_-, H)$ \cite{Gates:1984nk}.  
  
 The correspondence  works as follows:  Starting from  the Gates-Hull-Ro\v{c}ek geometry $({\cal M}, g, J_+, J_-, H)$
 we can construct  two $\tilde{H}$-twisted commuting generalized complex structures ${\cal J}_1$ and ${\cal J}_2$  
\beq
 {\cal J}_{1/2} = \frac{1}{2}  \left (\begin{array}{ll}
       1 & 0 \\
          b & 1 
       \end{array} \right )
\left ( \begin{array}{ll}
       J_+ \pm J_- & -(\omega_+^{-1} \mp \omega_-^{-1}) \\
     \omega_+ \mp \omega_- & - (J_+^t \pm J^t_-)
\end{array} \right )
\left (\begin{array}{ll}
       \,\,\,\,1 & 0 \\
         -b & 1 
       \end{array} \right )
\eeq{degindoftwocvom} 
 where $b$ is two-form (which maybe zero) and $H=\tilde{H}+db$.  The generalized metric $G=-{\cal J}_1 {\cal J}_2$ 
 is given 
\beq
 G = \left ( \begin{array}{ll}
-g^{-1}b  & g^{-1} \\
 g-bg^{-1}b & bg^{-1} 
\end{array} \right ) = \left (\begin{array}{ll}
       1 & 0 \\
          b & 1 
       \end{array} \right )
\left ( \begin{array}{ll}
 0 & g^{-1} \\
 g & 0 \end{array} \right )
\left (\begin{array}{ll}
       \,\,\,\,1 & 0 \\
         -b & 1 
       \end{array} \right ).
\eeq{definbigG}
 The opposite is also true: Starting from a (twisted) generalized K\"ahler geometry $({\cal M}, {\cal J}_1, {\cal J}_2)$  
 we can recover the complex structures $J_\pm$ with the right properties.

\section{D-branes in $O(d,d)$ formalism}
\label{s:N1branes}

 Before discussing D-branes (vs boundary conditions) for $N=(2,2)$ sigma model we consider the $N=1$ superconformal
 boundary conditions reformulated in terms of $T{\cal M}\oplus T^*{\cal M}$.  

 The boundary conditions for $N=(1,1)$ models have been discussed extensively during last years both at classical and
 quantum levels. Here we are interested in the geometrical interpretation of D-branes and our discussion is entirely 
 within the classical (semi-classical) sigma model. On the boundary the following gluing conditions are imposed
\beq
 \psi_-^\mu = \eta R^\mu_{\,\,\nu} (X)\psi_+^\nu ,
\eeq{fermansboudn}
 where $\eta = \pm 1$ corresponds to the choice of spin structure. 
  The expression (\ref{fermansboudn}) is most general local ansatz for the fermionic boundary conditions compatible 
 with dimensional analysis. The bosonic counterpart of (\ref{fermansboudn})
 and the restrictions on gluing matrix $R$ can be derived by requiring $N=1$ superconformal symmetry.
 This problem has been worked out in great details in \cite{Albertsson:2001dv, Albertsson:2002qc}, where the differential 
 conditions have been also considered. 

 Let us summarize the relevant results from \cite{Albertsson:2001dv, Albertsson:2002qc}. The gluing matrix 
 $R : T{\cal M} \rightarrow T{\cal M}$ has an interpretation in terms of submanifold with an extra data. 
  We defined a (maximal) projector $Q$ (i.e., $Q^2=Q$) such that $RQ=QR=-Q$. The complementary projector $\pi=
 1_d - Q$ defines the integrable distribution. Thus there is a maximal integral submanifold ${\cal D}$ which 
 corresponds to $\pi$. There exists a two-form $F\in \Omega^2({\cal D})$ on ${\cal D}$ such that $dF=H|_{\cal D}$. 
 Along ${\cal D}$ the gluing matrix is defined as follows
\beq
 \pi^t (g - F)\pi  = \pi^t (g + F)\pi R .
\eeq{definalobrla}
 As well the gluing matrix respects the metric, $R^t g R = g$ and therefore ${\cal D}$ is a Riemannian submanifold.
 For further use let us introduce the following object $r=\pi -Q$. It follows from above properties that $r^2=1_d$ 
 and $r^t gr =g$.   
 
 Now our goal is to reformulate above description in $O(d,d)$ covariant terms. 
Following the idea from \cite{Kapustin:2003sg} we can introduce 
\beq
  \psi^\mu = \frac{1}{2} (\eta\psi_+^\mu + \psi_-^\mu),\,\,\,\,\,\,\,\,\,\,\,\,\,\,
 \rho_\mu = \frac{1}{2} g_{\mu\nu} (\eta \psi_+^\nu - \psi_-^\nu)
\eeq{newfermcoor}
 where $\psi$ takes values in the pull-back of $T{\cal M}$ and $\rho$ in the pull-back of $T^*{\cal M}$.  
 Together they combine to ${\mathbf \Psi} = (\psi, \rho) \in T{\cal M} \oplus T^*{\cal M}$.
 We can introduce the new gluing matrix ${\cal R}$ defined on $T{\cal M} \oplus T^*{\cal M}$
\beq
{\cal R} = \left (\begin{array}{ll}
       1 & 0 \\
         F & 1 
       \end{array} \right )
\left ( \begin{array}{ll}
                  r &\,\,\,\,\, 0\\
                 0 & -r^t 
 \end{array}\right )
\left (\begin{array}{ll}
       \,\,\,\,\,1 & 0 \\
        - F & 1 
       \end{array} \right ) = \left (\begin{array}{ll}
                                    \,\,\,\,\,\,r & \,\,\,\,0 \\
                                   Fr + r^t F & -r^t 
                                 \end{array} \right ).
\eeq{simplerr}
The boundary conditions (\ref{fermansboudn}) are written now as follows 
\beq
{\cal R}{\mathbf \Psi}= {\mathbf \Psi} .
\eeq{newbouconds1}
In this boundary condition two form $F$ along ${\cal D}$ matters only. By explicit calculation one
 can show that the condition (\ref{newbouconds1}) with (\ref{simplerr}) encodes completely the algebraic part 
 of the boundary conditions (\ref{fermansboudn}). 

 Let us summarize the properties of ${\cal R}$ given by (\ref{simplerr}): ${\cal R}$ respects the natural pairing ${\cal I}$
 on $T{\cal M}\oplus T^*{\cal M}$, ${\cal R}^t {\cal I} {\cal R} = -{\cal I}$ (i.e., ${\cal R}\in so(d,d)$).
 Also ${\cal R}^2 = 1_{2d}$ and thus we can defined the projector operator $\frac{1}{2}(1_{2d} + {\cal R})$.
 This projector operator defines a real, maximal isotropic sub-bundle of $T{\cal M} \oplus T^*{\cal M}|_{\cal D}$:
\beq 
 \tau_{\cal D}^F =
\{ \frac{1}{2}(1_{2d} + {\cal R}) (X+\xi) = (X+\xi):\,\,\, X+\xi \in T{\cal D} \oplus T^*{\cal M}|_{\cal D},
\,\, \xi|_{\cal D} = i_X F \}.
\eeq{defintgebdks}
This coincides with the definition of generalized tangent bundle of generalized submanifold given in \cite{Gualtieri}.  
 Introducing the generalized metric $G$ as in (\ref{definbigG}) (but instead of $b$ use $F$) we can rewrite the property 
 $R^t g R = g$ as follows
\beq
 G {\cal R} +{\cal R} G =0 . 
\eeq{conformal1}
Indeed the differential conditions (i.e., the integrability of $\pi$ and $dF=H|_{\cal D}$)
  can be encoded in the requirement that 
 the sub-bundle (\ref{defintgebdks}) is involutive with respect to H-twisted Courant bracket
 defined on the sections of  $T{\cal M} \oplus T^*{\cal M}$, 
\beq
 (1_{2d} - {\cal R}) [ (1_{2d} + {\cal R})(X+\xi), (1_{2d} + {\cal R})(Y+\eta)]_H=0 .
\eeq{deficofk24}
Thus a D-brane (vs $N=1$ superconformal boundary condition) can be naturally described in terms of $T{\cal M}\oplus 
 T^*{\cal M}$. 

In this context we would like to make the following speculative observation. 
If ${\cal R}$ is lower triangular matrix with the properties ${\cal R}^t {\cal I} {\cal R} = - {\cal I}$ 
 and ${\cal R}^2=1_{2d}$ then it has a form (\ref{simplerr}) and we recover the standard world-sheet 
 description of D-brane. There is no need for the object ${\cal R}$ to be globally defined on whole ${\cal M}$.
  It is tempting to drop the condition for ${\cal R}$ to be a lower triangular. If we do this then
 D-brane corresponds to a (local) involutive maximal isotropic sub-bundle of $T{\cal M}\oplus T^*{\cal M}$. With this
 definition we lose a standard geometric interpretation of D-brane as a submanifold. 
 However due to the property (\ref{imprpsbcjna}) D-brane corresponds to 
  a foliation on ${\cal M}$ which maybe singular.
Infact at bosonic level this is allowed. Namely the conformality condition at the boundary is given by
\beq
 T_{++} - T_{--} = 4\d_1 X^\mu g_{\mu\nu} \d_0 X^\nu = 2\left (
                                                              \d_0 X\,\,\, g\d_1 X \right ) {\cal I} 
 \left (\begin{array}{l}
  \d_0 X \\ g\d_1 X \end{array}\right ) = 0
\eeq{condasbosc}
 and thus $(\d_0 X\,\, g\d_1 X )$ belongs to a maximal isotropic subspace of $T{\cal M}\oplus T^*{\cal M}$.
 These non-geometrical branes require more study and we hope to come back to the subject elsewhere. 

In next Section using the proposed description of $N=1$ branes we study $N=2$ branes. 
For the sake of clarity let us set $\eta=1$ for the rest of the paper. 

\section{Geometry of N=2 D-branes}
\label{s:N2branes}

 In this Section we describe D-branes for the general $N=(2,2)$ sigma model. 
 The formal description of these branes has been presented in \cite{Lindstrom:2002jb}. However the 
 geometrical interpretation has been lacking, except for K\"ahler case \cite{Ooguri:1996ck}. 
 Using the standard gluing matrix (\ref{fermansboudn}) the preservation of $N=2$ supersymmetry in 
 presence of the boundary implies
\beq
 J_- R = \pm R J_+
\eeq{definclass}
or alternatively
\beq
 R^t \omega_-  R = \pm \omega_+,
\eeq{defkaform}
 where $\pm$ correspond to different ways of gluing left and right $U(1)$ (super)currents.
 We call the case with $+$ a B-type and with $-$ an A-type branes. 
When $J_+ = J_-$ we recover the K\"ahler case with the standard terminology introduced in
\cite{Ooguri:1996ck}. 

Our goal is to use the observations from the previous Section and 
 reformulate the condition (\ref{definclass}) in terms of $T{\cal M} \oplus T^*{\cal M}$.
 Before doing this we can immediately observe one very important property of the branes:
they are coisotropic manifolds with respect to different Poisson structures.  
Namely from (\ref{defkaform}) it follows that $Q^t (\omega_+^{-1} \mp \omega^{-1}_-)Q=0$
 where $Q$ is a normal projector to a brane. Thus for B-type brane we have  
 $(\omega_+^{-1} - \omega^{-1}_-) N^* {\cal D} \subset T{\cal D}$, where $N^*{\cal D}$ is 
a conormal bundle of ${\cal D}$. Therefore the B-type branes are  coisotropic submanifolds
 with respect to Poisson structure $(\omega_+^{-1} - \omega^{-1}_-)$. Correspondingly the A-type 
 branes are coisotropic submanifolds with respect to  Poisson structure $(\omega_+^{-1} + \omega^{-1}_-)$.
 The distribution $(\omega_+^{-1} \mp \omega^{-1}_-) N^* {\cal D}$ on the coisotropic submanifold ${\cal D}$
 is involutive and the corresponding foliation is called the characteristic foliation. 
 It is interesting to point out that  the D-branes for the Poisson sigma model are 
 also coisotropic submanifolds with respect to a Poisson structure 
 \cite{Cattaneo:2003dp} (for related discussion see also \cite{Bonechi:2003hd}).

The conditions (\ref{definclass}) and (\ref{defkaform}) can be derived by looking at the appropriate gluing 
 conditions for the $U(1)$ currents
\beq
 j_+ \mp j_- = \psi_+^\mu \omega_{+\mu\nu} \psi_+^\nu \mp \psi_-^\mu \omega_{-\mu\nu} \psi_-^\nu = 0 .
\eeq{u1currenrs}
 It is useful to rewrite the linear combinations of $U(1)$ currents as follows 
\beq
 j_+ - j_- = 2{\mathbf \Psi}^t {\cal I} {\cal J}_{1} {\mathbf \Psi},\,\,\,\,\,\,\,\,\,\,\,\,\,
 j_+ + j_- = 2 {\mathbf \Psi}^t {\cal I} {\cal J}_{2} {\mathbf \Psi}
\eeq{newu1curtt}
 where ${\cal J}_1$ and ${\cal J}_2$ are $H$-twisted commuting generalized complex structures defined in 
 (\ref{degindoftwocvom}) with $b=0$. Combining the boundary condition (\ref{newbouconds1}) with 
 (\ref{u1currenrs})  and (\ref{newu1curtt}) we arrive on the following description of B- and A-types of 
 branes: For B-type brane ${\cal R}$ and ${\cal J}_1$ are commuting 
\beq
 {\cal R} {\cal J}_1 =  {\cal J}_1 {\cal R},
\eeq{conditionsforaaaB}
 and thus  sub-bundle $\tau_{\cal D}^{F}$ is stable under ${\cal J}_1$. 
 Therefore the sub-bundle $\tau_{\cal D}^F$ must decompose into $\pm i$ egeinbundles of ${\cal J}_1$
\beq
 \tau_{\cal D}^F = \tau_{\cal D}^{F+} \oplus \tau_{\cal D}^{F-}.
\eeq{decompossunkl}
 Since $\tau_{\cal D}^{F}$ is closed under the (twisted) Courant bracket (see the previous Section) then 
 each of $\tau_{\cal D}^{F\pm}$ is closed under the (twisted) Courant bracket. The distributions 
 $p(\tau_{\cal D}^{F\pm})$ are involutive and give rise to the corresponding foliations on ${\cal D}$.
 For A-type brane ${\cal R}$ and ${\cal J}_2$ are commuting 
\beq
{\cal R} {\cal J}_2 = {\cal J}_2 {\cal R}
\eeq{conditionsforaaaA}
 and thus sub-bundle $\tau_{\cal D}^{F}$ is stable under ${\cal J}_2$.
 Again as before $\tau_{\cal D}^F$ can be decomposed into $\pm i$ involutive eigenbundles of ${\cal J}_2$.
 Thus A- and B-types branes are both (twisted) generalized complex submanifolds 
 (as defined in \cite{Gualtieri})\footnote{The related discussion of submanifolds of generalized complex 
 manifolds is given in \cite{BB1}. However we follow the definitions from \cite{Gualtieri}.}
   however with respect to different  generalized complex structure. 

 Above we gave the geometric description of a generic B- and A-types branes for a general 
 $N=(2,2)$ sigma model. In the remaining part of this Section we discuss the special cases and 
 examples of these branes. Let us consider the case when $H$ is exact, $H=db$ and define the generalized
 K\"ahler geometry as in (\ref{degindoftwocvom}) and (\ref{definbigG}). Now we can consider a D-brane 
 ${\cal R}$ as defined in (\ref{simplerr}) with $F=b$. Combining the relations (\ref{conditionsforaaaB}), 
 (\ref{conditionsforaaaA})
 with (\ref{conformal1}) we end up with following description
\beq
  {\cal R} {\cal J}_1 = \pm {\cal J}_1 {\cal R},\,\,\,\,\,\,\,\,\,\,\,\,\,\,\,\,\,
 {\cal R} {\cal J}_2 = \mp {\cal J}_2 {\cal R}
\eeq{specacseofbr}
 correspondingly for B- and A-types of branes. These branes correspond to a special 
 case when  $J_-r = \pm r J_+$.

To illustrate the general discussion let us consider
  the well-known case of D-brane on a K\"ahler manifold \cite{Ooguri:1996ck}. 
 In this situation $J_+=J_-=J$ and the corresponding generalized complex structures are given by
\beq
 {\cal J}_1 = \left ( \begin{array}{ll}
                      J &\,\,\,\,\, 0 \\
                      0 & - J^t 
               \end{array}
\right ),\,\,\,\,\,\,\,\,\,\,\,\,\,\,\,\,\,\,\,
{\cal J}_2 = \left ( \begin{array}{ll}
                      0 & -\omega^{-1} \\
                      \omega  & \,\,\,\,\, 0
               \end{array}
\right ) .
\eeq{cyacase} 
 The B-type branes (\ref{conditionsforaaaB}) correspond to the K\"ahler submanifolds ${\cal D}$ with $F$ being closed 
 $(1,1)$ form with 
 respect to the complex structure $J$. The A-type branes (\ref{conditionsforaaaA}) correspond to the Lagrangian 
 submanifolds and ``coisotropic A-branes'' discovered in \cite{Kapustin:2001ij} (see also the related discussion in
  \cite{Lindstrom:2002jb}).  For the special case (\ref{specacseofbr}) B-branes correspond to
  the K\"ahler submanifolds with $F=0$ and A-branes to the Lagrangian submanifolds. 

The example with a non-trivial $H$ can be given by the group manifolds case. Let us consider a group manifold ${\cal G}$ with 
 a Lie algebra ${\mathbf g}$. If ${\mathbf g}$ admits a complex version of Manin triple
  $({\mathbf g}, {\mathbf g}_-, {\mathbf g_+})$ then  ${\cal G}$ has  a twisted generalized K\"ahler 
 geometry (vs $N=(2,2)$ WZW model). Namely a Lie algebra ${\mathbf g}$ should be equipped 
 with symmetric ad-invariant non-degenerate bilinear form $<\,\,,\,\,>$ and  ${\mathbf g}$ can be decomposed into a pair
 of maximally isotropic subalgebras ${\mathbf g}_-$ and ${\mathbf g}_+$ as a vector space. 
 The closed three form $H$ is given by $H(X,Y,Z)=<[X,Y],Z>$,
for further details of construction we refer to \cite{Lindstrom:2002vp}.
  In the same work \cite{Lindstrom:2002vp} the special class of 
 $N=2$ D-branes related to an automorphism of the affine algebras has been discussed. An example of  B-type 
 brane is given by the trivial Lie algebra automorphism. The underlying geometry of these branes has been studied in  
 \cite{Alekseev:1998mc} and a trivial automorphism corresponds to a conjugacy classes ${\cal D}$ on ${\cal G}$.
  Each conjugacy class admits a two-form $F$ such that $dF=H|_{\cal D}$. 
 Thus the conjugacy classes of ${\cal G}$ are the examples of twisted generalized complex submanifolds for ${\cal J}_1$,  
 the twisted generalized complex geometry. Using the generalization of these ideas  and 
 the notion of twisted conjuagacy classes \cite{Felder:1999ka, Stanciu:1999id}
 it is not hard to construct  more examples of the (twisted) generalized complex submanifolds. 

\section{Summary and discussion}
\label{s:end}

The geometry of the general $N=(2,2)$ sigma model is described in terms of two commuting 
 (twisted) generalized complex structures, ${\cal J}_1$ and ${\cal J}_2$. We have shown 
 that B-branes correspond to a (twisted) generalized complex submanifolds with respect to
 ${\cal J}_1$ and A-branes to a (twisted) generalized complex submanifolds with respect to 
 ${\cal J}_2$. The geometry of these branes in terms of tangent and cotangent bundles looks
 simple. However the intrinsic geometry of the branes can be quite involved (e.g., there are the foliations
 on the world-volume, see the previous Section) and presumably depends on the details of (twisted)
 generalized complex geometry. 
 Under the mirror symmetry automorphism of the world-sheet theory, the (twisted)
 generalized complex structures ${\cal J}_1$ and ${\cal J}_2$ get interchanged. Thus B- and
 A-type branes are interchanged under the mirror map as expected.    

The main lesson of this paper is that both $N=1$ and $N=2$ branes naturally described in 
 terms of $T{\cal M}\oplus T^*{\cal M}$. However in this language it is tempting to extend 
 the geometrical description of D-branes from a submanifold to a foliation (possibly singular).
 Next natural step is to construct the topological twists of the {\em general} $N=(2,2)$ sigma model.  
  Some important observations in this direction have been presented in \cite{Kapustin:2003sg}. 
 Hopefully better understanding of D-branes and topological twists of  the {\em general} $N=(2,2)$ sigma model
 will bring some light to a mirror symmetry beyond the Calabi-Yau case
(for the related discussion of mirror symmetry beyond Calabi-Yau manifolds
  see \cite{Gualtieri, BB2, Fidanza:2003zi}).
We plan to come back to this problem elsewhere.
\bigskip

\bigskip

{\bf Acknowledgements}:
 I am grateful to Marco Gualtieri and Alessandro Tomasiello for 
reading and commenting on the manuscript.

\end{document}